\documentstyle [prl,aps]{revtex}
\begin{document}

\draft
\title{High Temperature Symmetry Breaking, SUSY Flat Directions, 
and the Monopole Problem}

\author{
Gia Dvali$^{\P}$ and, Lawrence M.
Krauss$^{\S}$
}
\address{ 
$^{\P}$Physics Department, New York University, 4 Washington Place,
New York, NY 10003, and ICTP, Trieste, Italy\\
$^{\S}$Departments of Physics and Astronomy,
Case Western Reserve University,
10900 Euclid Ave, Cleveland OH 44106-7079
}
\wideabs{
\maketitle
\begin{abstract}
\widetext
Supersymmetric flat directions allow a generic counterexample to the
phenomenon of
symmetry restoration at high temperatures.
We show that (exponentially) large VEVs can be
developed along these directions through temperature-induced
`dimensional
transmutation'.
A minimum with broken gauge charges (e.g. electric
charge) can exist at arbitrarily high temperatures. In such a 
scenario
magnetic monopoles are never formed,
 independent of whether inflation occurs before
or after the GUT phase transition, or whether it occurs at all.
\end{abstract}

\pacs{}

}

\narrowtext

 It is well know that many supersymmetric theories and in particular the MSSM, contain
non-compact vacuum flat directions. These are certain trajectories in field space along
which the potential vanishes, no matter how large the expectation values of the
associated
scalar fields are. With unbroken supersymmetry these are flat to all orders in perturbation 
theory. $D$-flat directions can be parameterized with various holomorphic gauge invariants
constructed out of chiral superfields\cite{carlos}. Depending on the structure of the
superpotential, some of these invariants (or some combinations of them) can also drop-out of the
$F$-flatness conditions.  The space of all their possible values then will define
a continuous vacuum manifold, sometimes referred to as a moduli space. Each of the
independent holomorphic invariants accounts for a complex dimensionality one in the moduli
space. For example, consider a flat direction parametrized by an invariant constructed out
of $n$ chiral superfields $\phi_1\phi_1...\phi_n$. Far away from the origin, we can define
a canonically normalized (at the tree level) flat direction field $\phi = c
(\phi_1\phi_1...\phi_n)^{1/n}$ (where $c$ is a normalization constant). At any point
along the flat direction the masses of physical particles coupled to $\phi$
are $\sim \langle \phi \rangle$. Expanding about any of these vacua
$\phi \rightarrow \langle \phi \rangle + \phi$ we can ask what is the
effective low energy theory for the field $\phi$. By definition, $\phi$ is massless
and has no self-potential. The crucial point is that all the particles coupled to it 
(via
renormalizable couplings) 
get masses $\sim \langle \phi \rangle$ and can be integrated out. $\phi$
then can couple to the rest of the particles only through $\langle \phi
\rangle^{-1}$-suppressed
operators. Thus, at energies $<< \langle \phi \rangle$ the flat direction field is effectively
decoupled from the rest of the world! This simple fact will play a crucial role in what
follows.

 What is the thermal history of the flat directions? Since finite temperature effectively breaks
supersymmetry, the flat directions are lifted at high $T$. Usually it is assumed that at high
temperatures, the flat direction fields get confined to the origin where all the symmetries
are restored. This goes in accordance with a
general belief that spontaneously broken gauge (or global) symmetries must be restored at high
$T$.
In non-supersymmetric field theories counter-examples to this phenomenon are well
known\cite{nonrestore},
and it has also 
been shown that symmetry non-restoration takes place in various physically motivated 
minimal extensions
of the standard model\cite{DMS1}. The effect was also shown to persist 
by renormalization group-improved methods \cite{RG} and by lattice calculations
\cite{lattice}.

On the other hand, it is usually assumed that symmetry non-restoration is impossible
in supersymmetric theories. This may have to do with various 
no go theorems \cite{antinonrestore}
according to which in supersymmetric theories an interacting scalar
field always gets a $positive$ mass-square $\sim T^2$ and thus its expectation 
value will vanish. While this statement is correct for small field values, it does not
preclude the existence of other minima either for $\sim \langle \phi \rangle >> T$,
or
if the field in question has only non-renormalizable couplings\cite{DT}.\footnote{
In the present paper we will not be interested in scenarios which rely on a non-zero
external charge density\cite{charge}}
As it is argued below it turns out that such minima indeed exist in many cases of physical
and cosmological interest. Flat directions, in some sense, make the effect of
symmetry non-restoration more generic then in non-supersymmetric theories, where such flat
vacua are absent. In particular we show that high temperature
corrections can provide a runaway potential for the flat direction and induce
an {\it exponentially} large expectation value. Crudely the condition is that the gauge
couplings of the light fields that are in thermal equilibrium
should dominate over their Yukawa couplings to the heavy states that are getting
masses from the flat direction field. In realistic theories many flat
directions have this property and therefore the early history 
of the Universe may be very much different
from what one would otherwise naively guess. Our effect is reminescent of
Witten's upside-down hierarcy, in zero-temperature
supersymmetric theories where a
tiny supersymmetry breaking can induce an exponentially large VEV along the (would be)
flat direction provided certain conditions are met\cite{witten}.

The standard conclusion about the inevitability of symmetry restoration is based on the
following assumptions:

 1) large field values $\langle \phi \rangle >> T$ are
removed by some (zero temperature) potential, which grows
unbounded for these large  $\langle \phi \rangle$.

 2) The scalar field
in question $\phi$ is in thermal equilibrium.

Obviously one or both of these conditions are violated if $\phi$ is a
flat direction field.
As mentioned above, if one moves far enough ( so that $\langle \phi \rangle >> T$) 
along the flat direction,
all the
particles directly interacting with the field $\phi$ (call them $Q$-particles) become superheavy and $\phi$
decouples
from the rest of the light particle bath 
in thermal equilibrium (call them $q$-particles).
What is the impact of the thermal bath on $\langle
\phi \rangle$ in such a situation? Although light
$q$-particles, by
definition, are decoupled from $\langle \phi \rangle$ at the tree level, they
can couple to heavy $Q$-states. After integrating out these $Q$-particles, the parameters of
the effective low energy theory, e.g. such as masses and couplings of $q$-quanta,
will depend on the masses of the heavy states and thus on $\langle \phi \rangle$. 
In the other words the expectation value of the flat direction field
appears as the course graining scale for the low energy theory.
This induces a dependence of the free energy on $\langle \phi \rangle$. Depending on the
 precise nature
of the 
interaction, 
the resulting effective potential may force $\langle \phi \rangle >> T$. This is the 
central observation of the present paper.

We will now illustrate this general phenomenon in detail using several toy models. 

 Let $\phi$ be a flat direction field charged under some gauge symmetry, with
corresponding vector superfields $Q_a$. At the moment, we will assume that $\phi$
has no couplings in the superpotential. Let $q_{\alpha}$ be some light particles charged under the
same gauge symmetry. 
For large values of  $\langle \phi \rangle$ the gauge symmetry is broken
and $Q_a$ get masses $\sim \langle \phi \rangle$. The effective low energy theory
then consists of the massless $\phi$-quanta plus some light $q$-particles. Now, integrating out
the $Q_a$-states we get effective $\langle \phi \rangle$-suppressed non-renormalizable
couplings between $\phi$ and the $q$-states. These interactions appear in the K\"ahler
metric and at one loop they have a form
\begin{equation}
 K(\phi)_q^q qq^* =  qq^*\left ( 1 + {g^2C \over 16\pi^2}{\rm log}{\phi\phi^* \over M_P^2}\right )
\end{equation}
(the $\phi$-independent part is assumed to be rescaled).
Here $C$ is a (positive) Casimir operator 
and $M_P$ is the ultraviolate cut-off. The subscript denotes
a
derivative
with respect to the superfields. 
This result can be simply understood as the $\langle \phi \rangle$-dependent
wave function renormalization of the light fields. The low energy theories
at the different vacua of the flat direction are identical, modulo
masses of the integrated out heavy states. Thus these theories can be thought as one and the
same theory at different energy scales $\langle \phi \rangle$. Changing
$\langle \phi \rangle$ is then equivalent to changing the wave function renormalization.
Now let us consider this theory at a temperature $T$. We will always be interested in the
region $\langle \phi \rangle >> T$. For these values the contribution to the free
energy from the heavy gauge fields is exponentially suppressed $\sim T^4 {\rm e}^{- {\phi
\over T}}$ and can be neglected. The purely bosonic part of the action can then
be written as
\begin{equation}
 K(\phi)_q^q \partial_{\mu}q\partial ^{\mu}q^* -
K(\phi)_q^{q -1}|W_q|^2
\end{equation}
Qualitatively the generated runaway potential for the flat direction field can be understood as a
result of
the operators $\langle \partial^{\mu}q^*
\partial_{\mu}q \rangle$ and $\langle |W_q|^2 \rangle$ having nonzero thermal averages
at temperature $T$. The induced potential for $\phi$ favors larger 
$K(\phi)_q^q $ and thus larger $\phi$. In fact $\phi$ will grow at least until the
log becomes of the order of one, at which point the given approximation breaks down and
higher loop corrections must be included. Thus $\langle \phi \rangle$ will be
exponentially large
in this picture.

 To see more explicitly that the potential pushes $\phi$ to grow, we can 
expand the K\"ahler term
in terms of small perturbation of the absolute value around any point $|\phi| \rightarrow |\langle
\phi \rangle| + \rho$ and keep terms in the effective potential linear in $\rho$ 
(note that
$\rho$ may have either sign).
The constant part of the metric can be trivially absorbed in the wave function renormalization. The
resulting
non-derivative couplings of $\rho$ have the form:
\begin{equation}
({\rm negative~constant}){ \rho \over |\langle \phi \rangle|}  |W_q|^2  + {\rm higher~ terms}
\end{equation}
At high temperature this interaction generates a tadpole for $\rho$. To see this fix
$\rho$ as an external leg and let $q$-propagate in the loop. Since by assumption
$\langle q \rangle$ is zero, only loops with no external $q$-legs contribute. The result can be
extracted from the
computation done in \cite{goran}. Take for example $W = \lambda q^3/3$, then the $q$-independent
contribution is given by two-loop "butterfly" diagrams \cite{goran}
\begin{equation}
  ({\rm negative~constant}){ \rho \over |\langle \phi \rangle|} {\lambda^2 T^4 \over 48}
\end{equation}
Thus $\rho$ wants to grow positive at any point of the trajectory: there is a runaway potential for
$\phi$.

 Alternatively, the shape of the potential can be found via a 
straightforward calculation
\cite{BS} \footnote{Borut Bajc and Goran Senjanovic suggested this approach using
a simple example with non-renormalizable K\"ahler metric. 
We thank them for this suggestion.}.
Here we will generalise this approach and argue that our result can be simply
interpreted as
a $\langle \phi \rangle$-dependent renormalization of coupling constants of the fields in thermal
equilibrium. Treat $\phi$ as a constant background field and go to the canonical basis for the other
fields by rescaling the K\"ahler metric: $q \rightarrow \sqrt {K(\phi)_q^q} q$. In terms of these
fields
the interaction part in the potential now becomes:
\begin{equation}
\lambda(\phi)^2 |q|^4
\end{equation}  
where
\begin{equation}
\lambda(\phi)^2 = {\lambda^2 \over  K(\phi)_q^{q 3}}
\end{equation}
can be treated as a renormalized $\langle \phi \rangle$-dependent coupling constant.
Computing the leading diagrams without extra $q$-legs, we get the following effective potential
for $\phi$
\begin{equation}
\lambda(\phi)^2 {T^4 \over 48}
\end{equation}
This dependence creates an
effective potential for the background field $\phi$ even if its quanta are not in a thermal
equilibrium. The minimization of the free energy pushes 
$K(\phi)_q^q$ to be as large as possible.
In our case $K(\phi)_q^q$ is obtained by the one-loop integration of the heavy vector superfields
of mass $\sim  \langle \phi \rangle$ and it grows only if $\langle \phi \rangle$
grows. Thus the potential forces $\langle \phi \rangle$ to grow.

Up to now we were ignoring possible Yukawa couplings of $\phi$ and were assuming that
the only heavy particles contributing to the wave function renormalization of
the light $q$-quanta were gauge superfields. Let us consider what would be the effect of
heavy chiral superfields in the analogous situation. For this we have to assume that
$\phi$ couples to some particles in the superpotential, which also couple to $q$ and thus
renormalize its K\"ahler metric.
Let $Q$ be such a state and take the superpotential to have the following form:
\begin{equation}
W = h\phi Q^2 + yQ^2q + \lambda {q^3 \over 3}
\end{equation}
(where $h$ and $y$ are coupling constants).
Now when we go along the flat direction $\langle\phi\rangle \rightarrow \infty$, 
both the
gauge fields and also the $Q$- particles gain masses $\sim
\langle \phi \rangle$ and should be integrated out. However, in the
one-loop corrected K\"ahler for the low energy fields the latter particles' 
contribution
appears with an opposite sign:
\begin{equation}
 K_q^q = 1 + {g^2C - y^2C' \over 16\pi^2}{\rm log}{\phi\phi^* \over M_P^2}
\end{equation}
($C'$ is a positive number, and $g$ and $C$ are defined as before). 
This means that the slope of the high-$T$ effective potential
depends on the relative strength of the gauge and Yukawa couplings. $\phi$ will grow if the
gauge
coupling is strong enough. However, along the flat direction 
$g$ and $y$ must be understood as the running
coupling constants at scale $\langle\phi\rangle$. As a result the coefficient of the logarithm 
may change sign at some point $\langle \phi \rangle >> T$ where its overall magnitude
is still small enough so that perturbation theory can be trusted.
One may conclude then that that a stable minimum develops. The resulting value of $\phi$ is
in general exponentially large with respect to $T$. Thus exponentially large VEVs
can be induced at high $T$ by dimensional transmutation!

In general there can be other temperature-independent (but model-dependent) stabilising factors, like
soft
terms and or the $M_P$-suppressed interactions either in the K\"ahler metric 
or in the superpotential.
These require separate study in each case. Some of these 
issues will arise shortly
when we discuss cosmological implications.

The high $T$ symmetry breaking by the flat direction field can have two immediate
implications. First this can provide a desired initial condition for the Affleck-Dine baryogenesis
scenario\cite{AD}. More importantly perhaps it can cure\cite{DKL} the monopole
problem\cite{preskill}. In the present context this mechanism should be viewed as a variant of the canonical
solution via symmetry non-restoration\cite{DMS}, 
although the expectation values in the
present case are much larger then in the canonical non-restoration scenario.
( For alternative ideas for non-inflationary solutions of the monopole problem
see\cite{screening,sweeping}). 

The simplest specific example occurs when $\phi$ is an electrically
charged flat direction, but not the GUT adjoint (which is assumed not to have a
flat potential).
In this scenario (as in the standard one) the VEV of the GUT Higgs vanishes
above the temperature $\sim 10^{16}$GeV. However, some of the
gauge charges remain broken by the $\phi$ VEV.  
For instance, in the minimal SUSY $SU(5)$  $\phi$ can be a flat direction parameterized by the invariant
\begin{equation}
\phi^3 = \bar 5^{\alpha}10^{\beta}\bar 5^{\gamma}
\end{equation}
where the $10~{\rm and}~\bar 5$ are matter superfields and $\alpha,~\beta, ~\gamma$ are generation
indexes.
These break the $SU(5)$ gauge group down to $SU(3)_c$.
 
The resulting cosmological scenario is as follows:
as discussed 
above, at high temperature there is a huge potential barrier $\sim T^4$ separating
the broken and unbroken phases.  If the Universe `starts out' in the
phase with broken electric charge, it can easily persist there all the
way until the GUT phase transition that occurs around $T \sim M_{GUT}$.
At this moment the adjoint Higgs gets a VEV, but 
monopoles are never formed, since the Universe is superconducting and
magnetic fields are completely screened. Then the temperature
gradually drops and the barrier between the superconducting and the normal
phases disappears and the flat direction can relax to the origin
therefore restoring electromagnetism.
The important point is that the flat directions must be able to relax to
the origin. This is an obvious requirement and has to do nothing with the monopole
problem. 
In general,  this will be the case since the flat directions are
lifted by the supersymmetry breaking
soft term $m^2\phi\phi^*$ at zero temperature (usually of order
TeV). Obviously for the charged and colored superfields these soft terms must
be positive so that the flat directions are stabilized at the origin at zero
temperature. However, at high temperature the effect of the soft terms is
negligible and the potential is dominated by the thermal corrections
that
induce a large VEV for the field in question.
At what temperatures the non-thermal soft terms become important is, in
general, model dependent, but even in the most unoptimistic cases this
will not happen before the GUT phase transition.  Thus the desired
suppression of monopole-formation will persist.
In general, depending on the explicit scenario for supersymmetry breaking
the soft masses may have a nontrivial temperature and $\phi$- dependence.

For example, assume that, as in conventional scenarios, the supersymmetry breaking
takes place dynamically, due to instanton effects of some strongly coupled
gauge group. Gaugino condensation in the hidden sector is an example. 
Then the soft terms may vanish at temperatures above the supersymmetry-breaking
scale in the hidden sector ($T_h \sim 10^{11}$GeV or so for the gravity mediated scenarios)
provided the two sectors have roughly the same
temperatures. This is reasonable to expect if they were in equilibrium at some
$T >> T_h$. In such a case the gaugino condensate can vanish above the
deconfinement temperature in the hidden sector and the potential will be governed by
the thermal supersymmetry breaking effects as discussed above.
If the hidden sector was never in equilibrium, or if the supersymmetry-breaking in
this sector is not dynamical but hard, then the soft terms may remain intact
at arbitrarily high temperatures. Even in this most unnatural case the
zero temperature potential can take over only below $T\ \sim \sqrt{M_Pm}$.
The charge breaking minimum gets destabilized below this temperature and
$\phi$ starts coherent oscillations about the true minimum, just as in the
Affleck-Dine scenario.
Since this temperature is well below the GUT phase transition temperature,
monopoles are not
produced. 

The scenario 
we have outlined can work in the minimal supersymmetric $SU(5)$ model without postulating
any extra physics.  Whether or not it is responsible for resolving the monopole
problem, the generic phenomenon of symmetry non-restoration at high temperature in
supersymmetric models may be of some importance.

{\it Acknowledgements:} 
We thank B. Bajc and G. Senjanovic for very useful discussions. We also thank
A. Linde for valuable discussions on the different version of this
phenonmenon and G. Gabadadze and V. Rubakov for comments.
LMK's research is supported in part by the DOE. 
LMK would like to thank the theory group at CERN for
their hospitality during the early phases of this work.

\end{document}